\documentclass[conference]{IEEEtran}
\usepackage{epsfig,cite,bm,amsmath}
\def\sg{\mathop{\mbox{\tiny (SG-CSIT)}}}
\def\sl{\mathop{\mbox{\tiny (SL-CSIT)}}}
\def\dl{\mathop{\mbox{\tiny (TL-CSIT)}}}
\def\slr{\mathop{\mbox{\tiny (SL-CSIR)}}}
\def\dlr{\mathop{\mbox{\tiny (TL-CSIR)}}}
\def\dfr{\mathop{\mbox{\tiny (TG-CSIR)}}}
\def\df{\mathop{\mbox{\tiny (TG-CSIT)}}}
\IEEEoverridecommandlockouts

\begin{document}
\title{Distributed Opportunistic Scheduling for MIMO Ad-Hoc Networks}
\author{Man-On Pun, Weiyan Ge, Dong Zheng, Junshan Zhang and H. Vincent Poor
\thanks{Man-On Pun and H. Vincent Poor are with the Department of Electrical Engineering, Princeton University, Princeton, NJ 08544.}
\thanks{Weiyan Ge and Junshan Zhang are with the Department of Electrical Engineering, Arizona State University, Tempe, AZ 85287.}
\thanks{Dong Zheng is with NextWave Wireless Inc., San Diego, CA 92130.}
\thanks{This research was supported in part by the Croucher Foundation under a post-doctoral fellowship, in part by the U. S. National Science Foundation under Grants ANI-02-38550, ANI-03-38807, CNS-06-25637, and CNS-07-21820 and in part by Office of Naval Research through Grant
N00014-05-1-0636.}}
\maketitle
\setcounter{page}{1} \thispagestyle{empty}

\begin{abstract}
Distributed opportunistic scheduling (DOS) protocols are proposed for multiple-input multiple-output (MIMO) ad-hoc networks with contention-based medium access. The proposed scheduling protocols distinguish themselves from other existing works by their explicit design for system throughput improvement through exploiting spatial multiplexing and diversity in a {\em distributed} manner. As a result, multiple links can be scheduled to simultaneously transmit over the spatial channels formed by transmit/receiver antennas. Taking into account the tradeoff between feedback requirements and system throughput, we propose and compare protocols with different levels of feedback information. Furthermore, in contrast to the conventional random access protocols that ignore the physical channel conditions of contending links, the proposed protocols implement a pure threshold policy derived from optimal stopping theory, i.e. only links with threshold-exceeding channel conditions are allowed for data transmission. Simulation results confirm that the proposed protocols can achieve impressive throughput performance by exploiting spatial multiplexing and diversity.
\end{abstract}

\section{Introduction}
In a wireless ad-hoc network, multiple users communicate over wireless links by autonomously determining network organization, link scheduling, and routing. Similarly to other wireless systems, wireless ad-hoc networks encounter critical design challenges imposed by time-varying fading channels and co-channel interference. In the conventional random access protocols (e.g. CSMA), a link proceeds to transmit its data after a successful channel contention, regardless of its current channel conditions. This may cause system throughput degradation if the successful link is deep-faded. To circumvent this obstacle, a {\em distributed} opportunistic scheduling (DOS) scheme has been proposed for wireless ad-hoc networks in \cite{Zheng07a}. In contrast with the conventional random access protocol, DOS restricts data transmission to the successful links whose channel conditions exceed thresholds pre-designed using optimal stopping theory. For those successful links with channel conditions below the threshold, their data transmission opportunity is forgone, which allows other links to re-contend for the channel. This process continues until a successful channel contention is achieved by a link with good channel conditions. It has been demonstrated in \cite{Zheng07a} that
DOS can substantially outperform the conventional random access protocol in terms of system throughput. Despite its good performance, the scheduling scheme proposed in \cite{Zheng07a} is devised for single-antenna ad-hoc networks.

\begin{figure*}[htp]
\begin{center}
\includegraphics[scale=0.85]{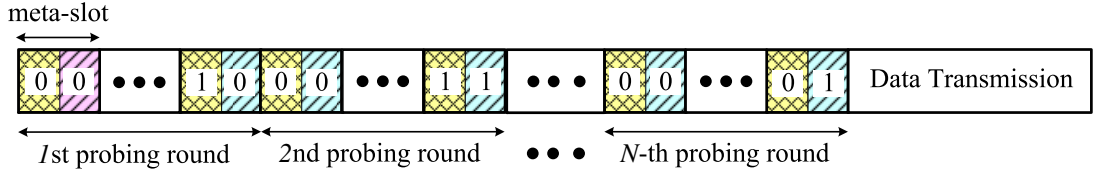}
\caption{An example of two-channel probing and data transmission where $0$ and $1$ indicate idle/collision state and successful channel contention, respectively.}\label{fig:model}
\end{center}
\end{figure*}

The recent success of multiple-input multiple-output (MIMO) techniques has inspired much research interest in MIMO ad-hoc networks. However, compared to the conventional single-link point-to-point MIMO transmission, by and large it is still an open issue on how to fully harvest the intrinsic spatial degrees of freedom of a MIMO ad-hoc network in a {\em distributed} fashion \cite{Hu04}. Taking an initial step, we propose novel MIMO DOS protocols with emphasis on exploiting spatial multiplexing and spatial diversity gains for MIMO ad-hoc networks in this work. More specifically, we consider a MIMO ad-hoc network where each user is equipped with $M$ antennas, where $M\geq 2$. To exploit the spatial multiplexing and diversity gains provided by the multiple antennas, we develop MIMO DOS protocols in which multiple links (in contrast to a single link in \cite{Zheng07a}) are probed and opportunistically selected. To facilitate the proposed multi-link channel contention and data transmission, a group-based splitting channel contention scheme is proposed. Based on the channel contention outcomes and the instantaneous rate, one or multiple links are scheduled to transmit over the $M$ spatial channels if their channel conditions exceed thresholds designed using optimal stopping theory. The resulting protocols are shown to achieve higher system performance by exploiting the spatial multiplexing and spatial diversity gains. Furthermore, we investigate the tradeoff between feedback requirements and throughput gain by developing protocols with either full channel state information (CSI) at the transmitter (CSIT) or CSI at the receiver only (CSIR). For the sake of presentational clarity, we concentrate on systems with $M=2$ in this work. However, it should be emphasized that the proposed protocol can be generalized for systems with $M>2$ antennas in a straightforward manner.


\section{Two-Group MIMO Scheduling Protocol with CSIT (TG-CSIT)}
\subsection{Protocol Description}
We consider a single-hop ad-hoc network in which each node is equipped with two antennas.  Suppose $K$ active links, {\em i.e.} $K$ pairs of source-destination (S-D) nodes $\left\{S_k,D_k\right\}$, $k=1,2,\cdots,K$, contend for data
transmission over two spatial channels constructed by transmit beamforming.

Each source node first randomly categorizes itself into one of the two equal-probable groups, namely Group 1 and Group 2, before its first channel contention. To facilitate the channel probing of each group, the channel time is divided into {\em meta-slots} composed of two {\em mini-slots} of duration $\tau$, each of which is exclusively assigned to one group as shown in Fig. \ref{fig:model}. A source node contends for both spatial channels in the mini-slots assigned to its group while all idle nodes eavesdrop communications of both groups. If only one link has contended in a mini-slot, the channel contention is considered successful under the assumption that both the designated destination node and other idle nodes can perfectly decode the contending messages. We define the random duration of achieving at least one successful channel contention in a meta-slot as one round of channel probing.

Let $i$ and $m$ denote the group and spatial channel indices with $i,m\in\left\{1,2\right\}$, respectively. For presentational convenience, we denote by $\left\{c_1,c_2\right\}$ the channel state of a meta-slot with $c_i$ being ``$1$" for successful channel contention by the $i$-th group and ``$0$" for either unsuccessful channel contention or lack of active links (idle). Thus, each probing round is completed with one of three possible channel states, namely, $\left\{0,1\right\}$, $\left\{1,0\right\}$ and $\left\{1,1\right\}$. At the end of each probing round, the destination node of each successful link returns information about the link conditions to its source node. Based on the feedback
information, the source node compares the channel conditions against thresholds pre-designed using optimal stopping theory, {\em i.e.} if the instantaneous transmission rate is above a pre-designed threshold, the source node will proceed data transmission; otherwise, the source node will forgo the opportunity and let other links to contend in the next meta-slot.

\subsection{Signal Model}
In this section we develop the signal model for channel state $\left\{1,1\right\}$, assuming $L_i$ is the successful link of the $i$-th group for $i=1,2$. Extension to the signal models for $\left\{0,1\right\}$ and $\left\{1,0\right\}$ will be outlined in Sec. \ref{sec:extension} whereas the trivial case $\left\{0,0\right\}$ is excluded from our following discussions.

For channel state $\left\{1,1\right\}$, $L_i$ is the only contending link in the $i$-th group. Thus, the received signal by node $D_{L_j}$ from node $S_{L_i}$, $j=1,2$, can be written as
\begin{equation}\label{eq:rlj}
{\bm y}_{L_i,L_j}=\sqrt{\rho_{L_i,L_j}}\cdot{\bm H}_{L_i,L_j}\cdot{\bm d}_{L_i}+{\bm n}_{L_j},
\end{equation}
where $\rho_{L_i,L_j}$ is the average signal-to-noise ratio (SNR), ${\bm H}_{L_i,L_j}$ is the complex channel gain matrix between node $S_{L_i}$ and node $D_{L_j}$, ${\bm d}_{L_i}$ is the transmitted data symbols from node $S_{L_i}$ with $E\left\{\left|{\bm d}_{L_i}\right|^2\right\}=1$ and ${\bm n}_{L_j}$ is modeled as a circularly symmetric white Gaussian noise with ${\cal CN}\left({\bm 0},{\bm I}\right)$. Note that  (\ref{eq:rlj}) represents the desired signal model for $i=j$ whereas it stands for the interference signal model for $i\neq j$.

In this work, we concentrate on a homogenous network and assume
\begin{equation}
\rho_{L_i,L_j}=\left\{\begin{array}{ll}\rho_s,&i=j,\\\rho_n,&i\neq j.\end{array}\right.
\end{equation}
Furthermore, we assume that node $D_{L_j}$ has perfect information about $\rho_{s}$, $\rho_{n}$ and ${\bm H}_{L_i,L_j}$ by exploiting the preambles transmitted from node $S_{L_i}$ during its channel contention.

\subsection{Statistics of Transmission Rate with CSIT}
Next, we investigate the statistics of transmission rate in the presence of two successful links for the following two cases: only one successful link is selected for data transmission or both links are scheduled to transmit data simultaneously.

\subsubsection{Single-link (SL) transmission}
We start from the case with only one transmission link, which amounts to the conventional MIMO point-to-point transmission. Assuming node $S_{L_i}$ has perfect information of ${\bm H}_{L_i,L_i}$ and constructs parallel transmission channels along the eigenvectors of ${\bm H}_{L_i,L_i}^H{\bm H}_{L_i,L_i}$, the resulting data rate is given by
\begin{equation}\label{eq:RLi}
r_{L_i}^{\sl}=\sum_{m=1}^2\log\left(1+\rho_s\lambda_{L_i,m}\right),
\end{equation}
where $\lambda_{L_i,m}$ is the eigenvalue of ${\bm H}_{L_i,L_i}^H{\bm H}_{L_i,L_i}$ for $m=1,2$. The joint probability density function (PDF) of the eigenvalues has been shown to be \cite{Telatar99}
\begin{equation}\label{eq:lambdajpdf}
f_{\bm\Lambda_{L_i}}\left(\lambda_{L_i,1},\lambda_{L_i,2}\right)=
\frac{1}{2}e^{-(\lambda_{L_i,1}+\lambda_{L_i,2})}\left(\lambda_{L_i,1}-\lambda_{L_i,2}\right)^2.
\end{equation}
Utilizing (\ref{eq:lambdajpdf}) and (\ref{eq:RLi}), the PDF of $R_{L_i}^{\sl}$ can be computed in the following fashion:
\begin{equation}\label{eq:PDFRL_i}
f_{R_{L_i}^{\sl}}\left(r\right)=\frac{1}{2}\int_{0}^{r}e^{-\left(\lambda_{L_i,1}+v\right)}
\left(\lambda_{L_i,1}-v\right)^2\,d\lambda_{L_i,1},
\end{equation}
where
\begin{equation}
v=\frac{1}{\rho_s}\left[2^{r-\log(1+\rho_s\lambda_{L_i,1})}-1\right].
\end{equation}

Subsequently, the corresponding CDF of $R_{L_i}^{\sl}$ can be obtained by
\begin{equation}\label{eq:CDFSL}
F_{R_{L_i}^{\sl}}\left(r\right)=\int_{0}^{r}f_{R_{L_i}^{\sl}}\left(r'\right)\,dr'.
\end{equation}

\subsubsection{Two-link (TL) transmission}
For the case in which both successful links transmit simultaneously, the signal received by node $D_{L_i}$ becomes the superposition of signals from the two source nodes and is given by
\begin{equation}\label{eq:rli2}
{\bm y}^{\dl}_{L_i}=\sqrt{\rho_{s}}\cdot{\bm H}_{L_i,L_i}\cdot{\bm d}_{L_i}+\sqrt{\rho_{n}}\cdot{\bm H}_{L_j,L_i}\cdot{\bm d}_{L_j}+{\bm n}_{L_i},
\end{equation}
where $j\neq i$. Clearly, (\ref{eq:rli2}) amounts to a point-to-point MIMO signal model in the presence of interference. Only a few analytical results on the exact channel capacity are available in the literature\cite{Win06}. To keep the following analysis tractable, we approximate the interference term as a zero-mean Gaussian noise with variance $\rho_n$. Thus, the data rate
of link $L_i$ is given by
\begin{equation}\label{eq:dlRLi}
r_{L_i}^{\dl}=\sum_{m=1}^2\log\left(1+\frac{\rho_s}{1+\rho_n}\lambda_{L_i,m}\right),
\end{equation}
and we have the sum-rate of the two links as
\begin{equation}\label{eq:dlRL1L2}
r_{L_1,L_2}^{\dl}=r_{L_1}^{\dl}+r_{L_2}^{\dl}.
\end{equation}

Recalling that ${\bm H}_{L_i,L_j}$ for $i,j\in\left\{1,2\right\}$ are statistically independent, the PDF of $R_{L_1,L_2}^{\dl}$ can be computed as
\begin{equation}
f_{R_{L_1,L_2}^{\dl}}\left(r\right)=\int_{0}^{r}f_{R_{L_1}^{\sl}}\left(r_{L_1}\right)
f_{R_{L_2}^{\sl}}\left(r-r_{L_1}\right)\,dr_{L_1},
\end{equation}
where $f_{R_{L_i}^{\sl}}\left(r_{L_i}\right)$ is given in (\ref{eq:PDFRL_i}) except that $\rho_s$ is replaced by $\frac{\rho_s}{1+\rho_n}$. Finally, the corresponding CDF can be obtained as
\begin{equation}\label{eq:CDFDL}
F_{R_{L_1,L_2}^{\dl}}\left(r\right)=\int_{0}^{r}f_{R_{L_1,L_2}^{\dl}}\left(r'\right)\,dr'.
\end{equation}

\subsection{Information Feedback}\label{sec:feeback}
\begin{figure}[htp]
\begin{center}
\includegraphics[scale=0.9]{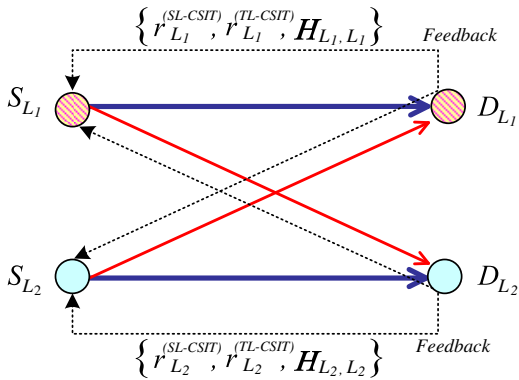}
\caption{Schematic diagram of the channel contention and feedback for TG-CSIT in channel state $\left\{1,1\right\}$.}\label{fig:feedback}
\end{center}
\end{figure}

Figure \ref{fig:feedback} gives a schematic diagram of the channel contention and feedback for channel state $\left\{1,1\right\}$. At the end of each probing round, node $D_{L_i}$ feeds back  $\left\{r_{L_i}^{\sl},r_{L_i}^{\dl},{\bm H}_{L_i,L_i}\right\}$ to node $S_{L_i}$.

Assuming that the feedback channel is error-free, node $S_{L_i}$ receives $\left\{r_{L_i}^{\sl}, r_{L_i}^{\dl}, {\bm
H}_{L_i,L_i}\right\}$ from node $D_{L_i}$ and overhears $\left\{r_{L_j}^{\sl}, r_{L_j}^{\dl}, {\bm H}_{L_j,L_j}\right\}$ from node $D_{L_j}$. Then, node $S_{L_i}$ will choose the transmission strategy that maximizes the transmission rate given by
\begin{equation}\label{eq:Rdf1}
r^{*\df}_{L_1,L_2}=\max\left\{r^{\sl}_{L_1},r^{\sl}_{L_2},r_{L_1,L_2}^{\dl}\right\}.
\end{equation}
For a reasonably large $\rho_s$ (more specifically, $\rho_s>\rho_n^2$), (\ref{eq:Rdf1}) can be simply approximated as
\begin{equation}\label{eq:Rdf}
r^{*\df}_{L_1,L_2}\approx r_{L_1,L_2}^{\dl},
\end{equation}
which holds when the multiplexing gain outweighs the signal-to-interference-noise (SINR) loss due to the presence of two simultaneously active links. In the following, we use (\ref{eq:Rdf}) in place of (\ref{eq:Rdf1}).

\subsection{Extension to Channel States $\left\{0,1\right\}$ and $\left\{1,0\right\}$}\label{sec:extension}

The signal model discussed above can be straightforwardly extended to channel states $\left\{0,1\right\}$ and $\left\{1,0\right\}$. Assuming $L_i$ is the only successful link, $r_{L_i}^{\sl}$ and $r_{L_i}^{\dl}$ can be evaluated from (\ref{eq:RLi}) and (\ref{eq:dlRLi}), respectively, except that $r_{L_i}^{\dl}$ contains no useful information in channel states $\left\{0,1\right\}$ and $\left\{1,0\right\}$. If node $S_{L_i}$ overhears {\em no} feedback beyond that from node $D_{L_i}$, node $S_{L_i}$ can safely assume that it is the only successful link and subsequently compare $r_{L_i}^{\sl}$ against the threshold for the data transmission decision.

\setcounter{equation}{18}
\begin{figure*}[htp]
\begin{equation}\label{eq:dfxopt}
x_{\max}^{\df}=\frac{\displaystyle\sum_{i=1,j\neq i}^2p_{i,s}\left(1-p_{j,s}\right)\int_{x_{\max}^{\df}}^\infty
r\,d\left[F_{R_{L_i}^{\sl}}(r)\right] +p_{1,s}\cdot
p_{2,s}\int_{x_{\max}^{\df}}^\infty
r\,d\left[F_{R_{L_1,L_2}^{\dl}}(r)\right]}
{2\delta+\displaystyle\sum_{i=1,j\neq i}^2p_{i,s}\left(1-p_{j,s}\right)\left(1-F_{R^{\sl}_{L_i}}(x)\right)
+p_{1,s}\cdot p_{2,s}\left(1-F_{R^{\dl}_{L_1,L_2}}(x)\right)},
\end{equation}\hrulefill
\end{figure*}
\setcounter{equation}{15}

\subsection{Optimal Threshold Design}
In this section, we invoke optimal stopping theory to design the rate threshold such that the decision whether to transmit data or not maximizes the average system throughput. Denote by $p_\ell$ and ${\cal I}_i$  the channel contention probability of the $\ell$-th link and  the link index set of Group $i$ for $i=1,2$, respectively.
The probability of a successful channel contention by links of Group $i$ can be computed as
\begin{equation}\label{eq:pis}
p_{i,s}=\sum_{\ell\in {\cal I}_i}p_\ell\prod_{\substack{j\in {\cal I}_i\\ j\neq \ell}} (1-p_j).
\end{equation}

Subsequently, the instantaneous transmission rate at the end of each probing round can be treated as a compound random variable (r.v.) as
\begin{equation}
R^{\df}=\underbrace{\displaystyle\sum_{i=1,j\neq i}^2p_{i,s}\left(1-p_{j,s}\right)\cdot R^{\sl}_{L_i}}_{\left\{1,0\right\} \mbox{ and } \left\{0,1\right\} }+\underbrace{p_{1,s}\cdot p_{2,s}\cdot R^{\dl}_{L_1,L_2}}_{\left\{1,1\right\}}.\label{eq:comprv}
\end{equation}

Invoking the renewal theorem, the rate of return after $N$ probing rounds can be subsequently defined as \cite{Zheng07a}
\begin{equation}
x^{\df}=\frac{E\left\{R^{\df}_{(N)}\cdot T\right\}}{E\left\{T_N\right\}},\label{eq:throughputgeneral1}
\end{equation}
where $T$ is the data transmission duration, and $T_N$ is the time duration including both $T$ and the time elapsed over the $N$ probing rounds. As shown in \cite{Zheng07a}, the optimal protocol that maximizes the average rate of return in
(\ref{eq:throughputgeneral1}) is a pure threshold policy and the maximal throughput $x^{\df}_{\max}$ can be found by solving (\ref{eq:dfxopt}) where $\delta=\tau/T$.

\subsection{Protocol Summary}
Here, we summarize the operation of the proposed protocol. Suppose the source node of a successful link, $S_{L_i}$, receives feedback on $\left\{r_{L_i}^{\sl},r_{L_i}^{\dl},{\bm H}_{L_i,L_i}\right\}$ from its destination node $D_{L_i}$ at the end of each probing round. If node $S_{L_i}$ cannot overhear feedback from the other group, it will compare $r_{L_i}^{\sl}$ against $x^{\df}_{\max}$ and proceed to data transmission only when $r_{L_i}^{\sl}\geq x^{\df}_{\max}$; otherwise,  $S_{L_i}$ will give up the transmission opportunity and let other links to contend for the channel in the next probing round. On the other hand, if $S_{L_i}$ detects the existence of another successful link from the
other group, it will compare $r_{L_i}^{\dl}+r_{L_j}^{\dl}$ against $x^{\df}_{\max}$, for $i\neq j$. Both links will transmit data simultaneously if $r_{L_i}^{\dl}+r_{L_j}^{\dl}$ exceeds $x^{\df}_{\max}$. Otherwise, no links will proceed to data transmission. During data transmission, ${\bm H}_{L_i,L_i}$ is exploited to generate the eigen-beamforming matrix necessary for creating the parallel channels.

In the sequel, this protocol is referred to as two-group MIMO scheduling with CSIT (TG-CSIT) since nodes are categorized into two groups for channel contention and perfect CSI is required at the transmit node.

\section{Two-Group MIMO Scheduling Protocol with CSIR (TG-CSIR)}
In TG-CSIT, an $M\times M$ channel gain matrix ${\bm H}_{L_i,L_i}$ is required to be returned to the source node of each successful link, which may entail a formidable feedback burden for a large $M$. To reduce the feedback amount, we consider a protocol that requires full CSI at the receivers only, and feedback of two real-valued data rates, which we refer to as the two-group MIMO scheduling protocol with CSIR (TG-CSIR) in the sequel.

Similar to TG-CSIT, TG-CSIR also splits the contending links into two groups and assigns mini-slots to each group for channel contention. As a result, each successful probing round also ends with one of the three states $\left\{1,0\right\}$, $\left\{0,1\right\}$ and $\left\{1,1\right\}$. However, TG-CSIR reduces the required feedback amount by allowing each link to transmit only one data stream over the two spatial channels. In other words, each link only reaps the diversity gain for states $\left\{1,0\right\}$ and $\left\{0,1\right\}$ whereas the spatial
multiplexing gain is exploited only in state $\left\{1,1\right\}$ by allowing two links to transmit simultaneously.

For states $\left\{1,0\right\}$ and $\left\{0,1\right\}$,  the received signal by node $D_{L_j}$ from node $S_{L_i}$ can be written as
\begin{equation}\label{eq:rlj-csir}
{\bm y}'^{\slr}_{L_i,L_j}=\sqrt{\rho_{L_i,L_j}}\cdot{\bm h}'_{L_i,L_j}\cdot{d}'_{L_i}+{\bm n}'_{L_j},
\end{equation}
where ${\bm h}'_{L_i,L_j}$ is the effective channel gain vector between nodes $S_{L_i}$ and $D_{L_j}$, ${d}'_{L_i}$ is the transmitted data symbol and ${\bm n}'_{L_j}$ is the noise term modeled as ${\cal CN}\left(0,{\bm I}\right)$. Since perfect CSI is assumed to be available to node $D_{L_i}$, the maximal SNR $\gamma^{\slr}_{L_i}$ is achieved by pre-multiplying the received signal with ${\bm h}'^H_{L_i,L_i}$. After some calculation, we can find the CDF of $\gamma^{\slr}_{L_i}$ as
\begin{equation}
F_{\Gamma^{\slr}_{L_i}}\left(\gamma\right)=1-\left(1+\frac{\gamma}{2}\right)e^{-\frac{\gamma}{2\rho_s}}.
\end{equation}
Thus, the resulting data rate is given by
\begin{equation}
r_{L_i}^{\slr}=\log(1+\rho_s\gamma_{L_i})
\end{equation}
whose CDF takes the following form.
\begin{equation}
F_{R^{\slr}_{L_i}}\left(r\right)=1-\left(1+\frac{2^r-1}{2\rho_s}\right)e^{-\frac{2^r-1}{2\rho_s^2}}.
\end{equation}

For state $\left\{1,1\right\}$, node $D_{L_i}$ is interfered by node $S_{L_j}$ for $i\neq j$ and its received signal can be written as
\begin{equation}\label{eq:rli2-csir}
{\bm y}'^{\dlr}_{L_i}=\sqrt{\rho_{s}}\cdot{\bm h}'_{L_i,L_i}\cdot{d}'_{L_i}+\sqrt{\rho_{n}}\cdot{\bm h}'_{L_j,L_i}\cdot{d}'_{L_j}+{\bm n}'_{L_i}.
\end{equation}
Since ${\bm h}'_{L_i,L_i}$ and ${\bm h}'_{L_j,L_i}$ are known to node $D_{L_i}$, the optimal combining (OC) technique provides the maximum SINR $\gamma^{\dlr}_{L_i}$ whose CDF can be shown to be \cite{Monzingo80,Villier99}
\begin{equation}
F_{\Gamma^{\dlr}_{L_i}}\left(\gamma\right)=1-\left(1+\frac{1}{2\rho_n}\right)e^{-\frac{\gamma}{\rho_s}}+
\frac{1}{2\rho_n}e^{-\frac{1+2\rho_n}{\rho_s}\gamma}.
\end{equation}
Hence, the CDF of the resulting data rate $r_{L_i}^{\dlr}=\log\left(1+\gamma^{\dlr}_{L_i}\right)$ is given by
\begin{equation}
F_{R^{\dlr}_{L_i}}\left(r\right)=1-\left(1+\frac{1}{2\rho_n}\right)e^{-\frac{2^r-1}{\rho_s}}+
\frac{1}{2\rho_n}e^{-\frac{1+2\rho_n}{\rho_s}\left(2^r-1\right)},
\end{equation}
and the corresponding PDF can be computed as
\begin{equation}
f_{R^{\dlr}_{L_i}}\left(r\right)=\left(1+\frac{1}{2\rho_n}\right)\frac{r2^{r-1}}{\rho_s}\left(e^{-\frac{2^r-1}{\rho_s}}
-e^{-\frac{1+2\rho_n}{\rho_s}\left(2^r-1\right)}\right).
\end{equation}
Finally, we are ready to compute the sum-rate of the two links defined as
\begin{equation}\label{eq:dlrRL1L2}
r_{L_1,L_2}^{\dlr}=r_{L_1}^{\dlr}+r_{L_2}^{\dlr}.
\end{equation}
Since ${\bm h}'_{L_i,L_j}$ for $i,j\in\left\{1,2\right\}$ are assumed to be statistically independent, $r_{L_1}^{\dlr}$ and $r_{L_2}^{\dlr}$ are also statistically independent. Therefore, the CDF of $R_{L_1,L_2}^{\dlr}$ is given by
\begin{equation}\label{eq:CDFDLr}
F_{R_{L_1,L_2}^{\dlr}}\left(r\right)=\int_{0}^{r}
F_{R^{\dlr}_{L_2}}\left(r-r_{L_1}\right)f_{R^{\dlr}_{L_1}}\left(r_{L_1}\right)\,dr_{L_1}.
\end{equation}

Upon obtaining $F_{R^{\dlr}_{L_i}}\left(r\right)$ and $F_{R_{L_1,L_2}^{\dlr}}\left(r\right)$, we can treat the instantaneous data rate as a compound r.v. $R^{\dfr}$ and define the corresponding rate of return $x^{\dfr}$ in a similar fashion as shown in (\ref{eq:comprv}) and (\ref{eq:throughputgeneral1}). It is clear that the optimal strategy for TG-CSIR is also a pure threshold policy in which the threshold can be computed from (\ref{eq:dfxopt}) with $F_{R^{\dl}_{L_i}}\left(r\right)$ and $F_{R_{L_1,L_2}^{\dl}}\left(r\right)$ replaced by $F_{R^{\dlr}_{L_i}}\left(r\right)$ and $F_{R_{L_1,L_2}^{\dlr}}\left(r\right)$, respectively. It is easy to see that TG-CSIR requires only node $D_{L_i}$ to feed back $\left\{r_{L_i}^{\slr},r_{L_i}^{\dlr}\right\}$, irrespective of $M$, which results in substantial feedback reduction for a large $M$ compared to TG-CSIT. However, this feedback reduction is obtained by sacrificing partial achievable multiplexing gain.

\section{Single-Group MIMO Scheduling (SG) Protocol with CSIT (SG-CSIT)}
For comparison purposes, we also consider a straightforward extension of the DOS scheme proposed in \cite{Zheng07a}. In this protocol, all links are allowed to contend in each mini-slot and the transmission rate of a successful link, $r^{\sg}$ , is given by (\ref{eq:RLi}). Since all nodes  belong to one group compared to two groups in TG-CSIT/CSIR, this protocol is referred to as single-group MIMO scheduling protocol with CSIT (SG-CSIT) in the sequel. It is straightforward to show that the average rate of return of SG is given by
\begin{equation}\label{eq:throughputgeneral3}
x^{\sg}=p'_{s}\frac{E\left[r^{\sg}T\right]}{E\left[T_{N}\right]},
\end{equation}
where $p'_{s}$ is defined in (\ref{eq:pis}) with ${\cal I}_i$ containing all link indices.

The corresponding maximal average rate of return can be found by solving (\ref{eq:sgxopt}):
\begin{equation}\label{eq:sgxopt}
x_{\max}^{\sg}=\frac{p'_s\int_{x_{\max}^{\sg}}^\infty
r\,d\left[F_{R^{\sg}}(r)\right]}{\delta+p'_{s}\left[1-F_{R^{\sg}}(x_{\max}^{\sg})\right]},
\end{equation}
where $F_{R^{\sg}}(r)$ can be computed from (\ref{eq:CDFSL}).

\section{Simulation Results}

In this section, we conduct Monte Carlo experiments to assess the performance of the proposed protocols. Unless otherwise specified, we set $p_{1,s}=p_{2,s}=e^{-1}$, $\rho_n=1$ and $\delta=0.1$. Furthermore, the unit of throughput is nats/sec/Hz.

\begin{figure}[htp]
\begin{center}
\includegraphics[scale=0.33]{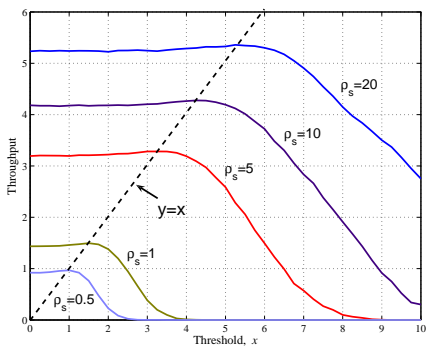}
\caption{Throughput as a function of the threshold with different $\rho_s$ values for TG-CSIT.}\label{fig:throughput_threshold}
\end{center}
\end{figure}

Figure \ref{fig:throughput_threshold} shows the network throughput as a function of the threshold for TG-CSIT. It is clear from Fig. \ref{fig:throughput_threshold} that the throughput is maximized when the threshold is set to $x_{\max}^{\df}$, which is indicated by the line labeled ``$y=x$".

\begin{figure}[htp]
\begin{center}
\includegraphics[scale=0.33]{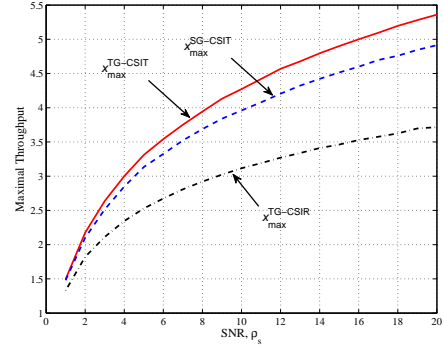}
\caption{Maximal throughput of the proposed protocols as a function of SNR.}\label{fig:different_scheme}
\end{center}
\end{figure}

Figure \ref{fig:different_scheme} depicts the maximal throughput as a function of $\rho_s$ for the proposed schemes. Inspection of Fig. \ref{fig:different_scheme} confirms that TG-CSIT achieves substantial performance improvement over SG-CSIT. In particular, Fig. \ref{fig:different_scheme} reveals that TG-CSIT outperforms SG-CSIT and TG-CSIR by $10\%$ and $40\%$ at SNR=$20$ dB, respectively. It should be emphasized that SG-CSIT achieves more throughput over TG-CSIR at the cost of more feedback amount.

\section{Conclusion}
In this paper, three distributed opportunistic scheduling protocols, namely TG-CSIT, TG-CSIR and SG-CSIT, have been proposed for MIMO ad-hoc networks. To fully harvest the multiplexing and spatial diversity gains provided by multiple antennas in a distributed fashion, links are divided into groups during channel contention such that multiple links may be considered for data transmission simultaneously over parallel spatial channels generated by multiple transmit/receive antennas. Furthermore, to maximize the overall system throughput, a pure threshold policy has been derived using optimal stopping theory. Thus, data transmission is only scheduled for successful links with channel conditions exceeding the pre-designed threshold. It has been demonstrated by simulation that all three proposed protocols can achieve impressive throughput performance and that TG-CSIT outperforms TG-CSIR at the cost of increased feedback.

\bibliographystyle{IEEEtranS}
\bibliography{Bib}
\end{document}